## Journal of Geophysical Research: Space Physics

# The impact of energetic electron precipitation on mesospheric hydroxyl during a year of solar minimum

**Annet Eva Zawedde[1], Hilde Nesse Tyssøy[1], Robert Hibbins[1,2], Patrick J. Espy[1,2], Linn-Kristine Glesnes Ødegaard[1], Marit Irene Sandanger[1], and Johan Stadsnes[1]**

[1]Birkeland Centre for Space Science, Department of Physics and Technology, University of Bergen, Bergen, Norway,
[2]Department of Physics, Norwegian University of Science and Technology, Trondheim, Norway

**Abstract** In 2008 a sequence of geomagnetic storms occurred triggered by high-speed solar wind streams from coronal holes. Improved estimates of precipitating fluxes of energetic electrons are derived from measurements on board the NOAA/POES 18 satellite using a new analysis technique. These fluxes are used to quantify the direct impact of energetic electron precipitation (EEP) during solar minimum on middle atmospheric hydroxyl (OH) measured from the Aura satellite. During winter, localized longitudinal density enhancements in the OH are observed over northern Russia and North America at corrected geomagnetic latitudes poleward of 55°. Although the northern Russia OH enhancement is closely associated with increased EEP at these longitudes, the strength and location of the North America enhancement appear to be unrelated to EEP. This OH density enhancement is likely due to vertical motion induced by atmospheric wave dynamics that transports air rich in atomic oxygen and atomic hydrogen downward into the middle atmosphere, where it plays a role in the formation of OH. In the Southern Hemisphere, localized enhancements of the OH density over West Antarctica can be explained by a combination of enhanced EEP due to the local minimum in Earth's magnetic field strength and atmospheric dynamics. Our findings suggest that even during solar minimum, there is substantial EEP-driven OH production. However, to quantify this effect, a detailed knowledge of where and when the precipitation occurs is required in the context of the background atmospheric dynamics.

## 1. Introduction

Energetic particles precipitating into the mesosphere and lower thermosphere are known to produce copi-ous amounts of odd nitrogen ($NO_X$: N, NO, $NO_2$) and odd hydrogen ($HO_X$: H, HO, $HO_2$), which can contribute to ozone ($O_3$) destruction [e.g., *Jackman et al.*, 2005; *Sinnhuber et al.*, 2012]. The energetic particles (electrons, protons, and heavier ions) have different solar drivers. Coronal Mass Ejections (CMEs) associated with sunspots predominantly occur during solar maximum and are the cause of solar proton events (SPEs) which can lead to strong geomagnetic activity. The influence of the infrequent SPEs upon the middle atmosphere has been extensively studied [see, e.g., *Bates and Nicolet*, 1950; *Weeks et al.*, 1972; *Swider and Keneshea*, 1973; *Crutzen and Solomon*, 1980; *Solomon et al.*, 1981; *López-Puertas et al.*, 2005; *Damiani et al.*, 2008, 2010; *Verronen and Lehmann*, 2013; *Jackman et al.*, 2014; *Nesse Tyssøy and Stadsnes*, 2015]. The atmospheric effects of the more frequent energetic electron precipitation (EEP) events are less known and harder to detect. During geomag-netic storms energetic electrons are injected and stored in the magnetosphere where they can be accelerated to relativistic energies [*Foster et al.*, 2014] and subsequently lost to the atmosphere [*Turner et al.*, 2014]. The penetration depth varies with the particle energy, for example, a 30 keV electron will stop at ∼90 km, while a 1 MeV electron penetrates to about 60 km [*Turunen et al.*, 2009]. Individually, such storms have weaker geomagnetic signatures than SPEs. It is, however, speculated that these events, because of their frequent occurrence, will have a strong impact on the atmosphere in general [*Andersson et al.*, 2014a].

*Bartels* [1932] identified "*M* regions" on the solar surface as the source of the sequences of recurrent geo-magnetic activity that occurred during minimum solar activity. *M* regions are in fact coronal holes (CHs) and are independent of sunspot activity [*Allen*, 1943]. They are associated with open magnetic field lines and, high-speed, low-density flows in the solar wind [*Billings and Roberts*, 1964]. CHs are the source of high-speed solar wind streams (HSSWS) and subsequent recurrent geomagnetic activity [e.g., *Neupert and Pizzo*, 1974; *Burlaga and Lepping*, 1977; *Sheeley and Harvey*, 1981]. The interaction of the fast solar wind associated with CHs with the slow solar wind streams results in the compression of the magnetic field and plasma at their interfaces forming a corotating interaction region (CIR), which is the geoeffective structure [*Tsurutani et al.*, 2006; *Gopalswamy*, 2008]. However, the interplanetary magnetic field (IMF) associated with CIRs has a highly oscillating nature, which results in only moderate intensification of the magnetospheric currents and hence moderate geomagnetic signatures. The intensity of the resulting storm depends on the combination of solar wind speed and the direction of the $B_z$ component [*Gopalswamy*, 2008].

Recent studies [*Verronen et al.*, 2011; *Andersson et al.*, 2012, 2014b, 2014a] provide observational evidence of radiation belt (geomagnetic latitudes 55°–65°) electron precipitation (100–300 keV) affecting mesospheric (71–78 km) OH. Based on two case studies in the declining phase of the solar cycle, *Verronen et al.* [2011] found that 56–87% of the changes in OH could be explained by changes in EEP. In a follow up study, *Andersson et al.* [2012] focused on a larger part of the solar cycle from solar maximum to solar minimum. They found months of high correlation between daily zonal mean OH mixing ratios at 70–78 km and the flux of 100–300 keV electrons. The correlation coefficients were highly dependent on season and the strength of the particle precipitation. *Andersson et al.* [2014b] studied the longitudinal response of nighttime mesospheric OH to >30 keV electron precipitation, contrasting days with daily mean count rates of >100 c/s to days with <5 c/s. In total 51 days between 2005 and 2009 met the first criteria. Generally, they concluded that clear effects of EEP were seen at magnetic latitudes 55°–72°. In the Southern Hemisphere (SH), the OH data revealed localized OH mixing ratio enhancements at longitudes between 150°W and 30°E, over West Antarctica, poleward of the South Atlantic Magnetic Anomaly (SAMA) region. In the Northern Hemisphere (NH), EEP-induced OH variations were more equally distributed with longitude; however, two potential regions of enhanced OH mixing ratio above Northern America and Northern Russia were found.

The middle atmosphere has a strong seasonal dynamical variability, including both the background meridional and zonal winds, as well as the atmospheric wave activity [see, e.g., *Shepherd*, 2000; *Kleinknecht et al.*, 2014]. For example, *Damiani et al.* [2010] have shown that during sudden stratospheric warmings (SSWs), the OH layer may show short-term variations comparable in strength to the OH increases during SPEs. *Andersson et al.* [2014b] did not include potential seasonal or meteorological factors when considering the particle impact upon the longitudinal distribution of OH, although there appears to be features less constrained to the magnetic latitudes and geomagnetic activity in both hemispheres. During strong particle precipitation events, the OH production due to background dynamics of the atmosphere might be overshadowed by the impact of energetic particle precipitation (EPP). However, during the more frequent and modest changes, the dynamical background will be of higher importance. Moreover, for the more frequent events, the magnitude of the direct EEP-induced $HO_x$ effect on $O_3$ in the mesosphere is high enough to suspect that EEP could be an important contribution to the Sun-climate connection on solar cycle time scales [*Andersson et al.*, 2014a]. Assessing the impact and spatial distribution of electron forcing is, therefore, important for more accurate modeling of its atmospheric and climate effects.

The quantification of relativistic electron precipitation has, however, proved difficult due to particle detector challenges [see, e.g., *Nesse Tyssøy et al.*, 2016]. In addition, radiation belt electrons usually have strong anisotropic pitch angle distribution that needs to be accounted for when considering their impact upon the atmosphere [*Rodger et al.*, 2013; *Nesse Tyssøy et al.*, 2016]. In this study, we optimize the data from the Medium Energy Proton and Electron Detectors (MEPED) on the Polar Orbiting Environmental Satellite (POES) NOAA-18, taking into account detector degradation, proton contamination, and combining data from both the 0° and 90° telescopes to achieve a better estimate of the true loss cone fluxes. We also use electron fluxes with energy >1000 keV obtained from the proton telescopes to determine the EEP impact on OH in the middle atmosphere [*Nesse Tyssøy et al.*, 2016]. Whereas *Andersson et al.* [2014b] used all available POES satellites, we only use NOAA-18, which is traversing the same local time as the Aura satellite making it possible to study the local effects of the energy deposition by relativistic electrons on OH. The data and its application are further explained in the next section.

Since most studies have focused on geomagnetic activity during solar maximum, it is paramount to get a deeper understanding of the contribution of EEP on $HO_x$ also during solar minimum. Therefore, we target the solar minimum year of 2008, where a sequence of weak to moderate storms triggered by HSSWS occurred. The low intensity of the recurrent storms implies that we need to carefully consider the role of the changing background dynamics upon the OH distribution. In addition to OH mixing ratios, the Aura MLS provides measurements of temperature, water vapor ($H_2O$), and geopotential height (GPH) which reveal the background state of the atmosphere. Thus, extracting information on both the longitude and altitude distribution enables us for the first time, to separate the OH variability caused by EEP and by atmospheric dynamics. The resulting analysis is given in section 3, and the subsequent discussions and conclusion follows in sections 4 and 5.

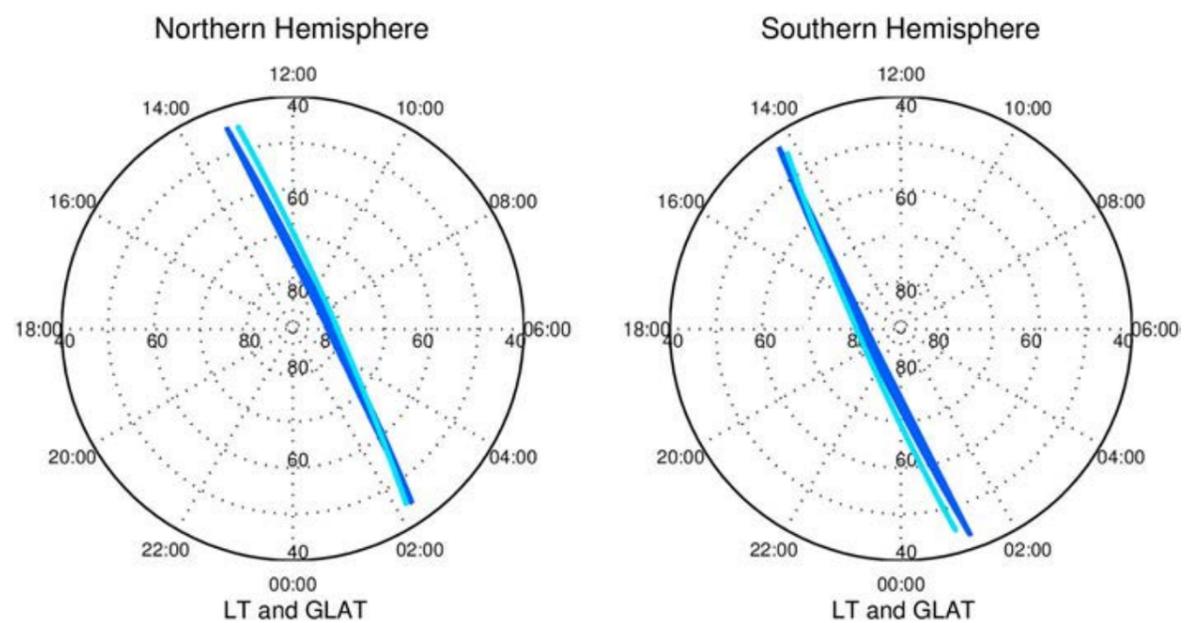

**Figure 1.** A plot showing the footprints of Aura (dark blue) and NOAA 18 (light blue) in local time and geographical latitudes for year 2008. Midnight is at the bottom and dusk to the left of each plot.

## 2. Materials and Methods

### 2.1. Aura MLS Observations

The Microwave Limb Sounder (MLS) is one of the four instruments on board NASA's Aura satellite [*Schoeberl et al.*, 2006]. It is in a near-polar Sun-synchronous orbit at 705 km altitude, scanning the atmosphere up to geographic latitudes 82° N/S with about 14 orbits per day (period ∼100 min). The MLS measures naturally occurring microwave thermal emissions from the limb of the Earth's atmosphere to remotely sense vertical profiles of atmospheric constituents [*Schoeberl et al.*, 2006; *Waters et al.*, 2006].

In this study we use Aura/MLS level 2 files version 4.2x for the year 2008 screened as per *Livesey et al.* [2015]. Only nighttime observations with solar zenith angle (SZA) >100° are considered to make sure no sunlight illuminates the sampled atmosphere below 100 km [*Pickett et al.*, 2006]. At night without solar radiation, the data should typically show low values of background OH, which makes the detection of OH enhancements due to EEP effects easier. In the latitude range of the radiation belts, nighttime Aura measurements occur at local solar time (LST) 02:15-03:30 in the NH during 2008. In the SH, however, Aura measurements are from LST 15:26 to 01:18. For SH nighttime observations we use LST 22:00-01:18.

The temporal resolution of Aura MLS data is ∼25 s. The vertical and horizontal resolution of OH measurements is 2.5 km and 165 km, respectively, within mesospheric altitudes (60–80 km). $O_3$, $H_2O$, temperature, and GPH have coarser and variable vertical/horizontal resolutions within mesospheric altitudes [see *Livesey et al.*, 2015].

The geometric height, $z$, can be expressed using the pressure altitude as

$$z = -H \ln \left( \frac{P}{P_s} \right) \qquad (1)$$

where $H$ is the atmospheric scale height (∼7 km) [*Brasseur and Solomon*, 2005], $P_s$ is a reference pressure (1000 hPa) and $P$ is the pressure level given in the MLS data.

### 2.2. NOAA POES MEPED Observations

In 2008, one of the five NOAA/POES satellites, NOAA 18 scanned the Earth at approximately the same local times as the Aura satellite (see Figure 1). This implies that the particle fluxes measured by MEPED/NOAA 18 deposited their energy close in both time and space to the measurements performed by Aura. Considering the short lifetime of OH below 80 km, the EEP impact on OH is considered to be a local effect.

The MEPED consists of two proton and two electron telescopes viewing almost perpendicular to each other. The electron and proton telescopes pointing radially outward are often named the 0° detectors. At high latitudes they will view particles within the loss cone and have therefore previously been used to represent the precipitating fluxes [*Verronen et al.*, 2011; *Andersson et al.*, 2012, 2014b, 2014a]. Electrons with energies capable of precipitating into the middle atmosphere (>30 keV) have, however, often a strongly anisotropic pitch angle distribution, which decreases toward the center of the loss cone. The 0° detector looking close to the center of the loss cone will therefore provide an underestimate of the precipitating fluxes. The other electron and proton telescope, called the 90° detector, view particles near the edge or outside the loss cone. Therefore, the 90° detectors will measure higher fluxes compared to the true precipitating fluxes. To overcome this challenge, we combine data from both electron telescopes to estimate the electron fluxes over the entire loss cone. The 0° and 90° electron fluxes were fitted to the solution of the Fokker-Planck equation for pitch angle diffusion of energetic particles [*Kennel and Petschek*, 1966]. We take into account the detector sensitivity when the directional flux varies over the acceptance solid angle of the telescope [*Nesse Tyssøy et al.*, 2016]. We correct the electron data for contamination by protons [*Yando et al.*, 2011]. The degradation of the proton detector is taken into account by applying the new correction factors developed by *Sandanger et al.* [2015].

**Table 1.** MEPED Proton and Electron Energy Channels [*Evans and Greer*, 2000]

| Proton Energy Channels | | Electron Energy Channels | |
|---|---|---|---|
| Channel | Energy Range (keV) | Channel | Energy Range (keV) |
| 0/90 P1 | 30 to 80 | 0/90 E1 | 30 to 2500 |
| 0/90 P2 | 80 to 240 | 0/90 E2 | 100 to 2500 |
| 0/90 P3 | 240 to 800 | 0/90 E3 | 300 to 2500 |
| 0/90 P4 | 800 to 2500 | | |
| 0/90 P5 | 2500 to 6900 | | |
| 0/90 P6 | >6900 | | |

During solar minimum, when no SPEs occur, there will be insignificant high-energy proton fluxes detected by the MEPED proton telescope which can be confirmed by the P4 and P5 energy channels (see Table 1). Considering that the highest-energy channel of the proton detector (P6) is responsive to relativistic electrons, we utilize this contamination effect to get a quantitative measure of electron fluxes larger than 1 MeV [*Yando et al.*, 2011]. The particle fluxes are sampled every 2 s. However, for purposes of comparing the particle fluxes with composition data from Aura, we average the fluxes over 1° latitude bin equivalent to about 16 s.

Combining the electron and proton channels (E1, E2, E3, and P6), we achieve a differential electron spectrum covering energies from 50 to 1000 keV. We use the electron spectra to calculate the energy deposition as a function of altitude. In these calculations, we use the cosine-dependent Isotropic over the Downward Hemisphere (IDH) model of *Rees* [1989]. This is a range-energy analysis based on a standard reference atmosphere (COmmittee on SPAce Research International Reference Atmosphere 1986).

### 2.3. OMNI Data
In this study we use IMF and solar wind plasma parameter data for the year 2008 downloaded from the Coordinated Data analysis Web (http://cdaweb.gsfc.nasa.gov./istp-public/). We focus on the $B_z$, $V_{SW}$, and solar wind flow pressure ($P_{SW}$), at 1 h resolution in geocentric solar magnetospheric (GSM) coordinates. Originally, the data are measured by either WIND or the ACE satellite and time shifted to the Earth's bow shock nose. The *Dst* and *AE* indices also have a 1 h resolution, while the *Kp* index has a 3 h resolution.

## 3. Results
### 3.1. Solar Wind and Geomagnetic Conditions
In Figure 2, daily means of *Kp*, *Dst*, *AE*, $B_z$, $V_{SW}$, and $P_{SW}$ are plotted for the whole year 2008, during which two HSSWS associated with the 27 days solar rotation period are indicated by the vertical dashed and dotted lines. The daily mean *Dst* index and IMF $B_z$ clearly show decreases corresponding to the arrival of the HSSWS, indicating the recurrent storm activity experienced by the magnetosphere. All the storms were weak to moderate based on the *Dst* index classification by *Loewe and Prölss* [1997].

In the period January to April, both the HSSWS were geoeffective. In May, the signature of the HSSWS in the geomagnetic indices is weak. In the daily mean *Dst*, for example, this period shows *Dst* > −15 nT. Both *Kp* and *AE* also display low values in May. From June to September, one of the HSSWS becomes geoeffective again (dashed vertical line). The signature from this HSSWS disappears in October. The other HSSWS (dotted vertical line) also becomes geoeffective again in September. At the end of the year, in November and December, the solar wind and geomagnetic signatures from the two HSSWS are weakened.

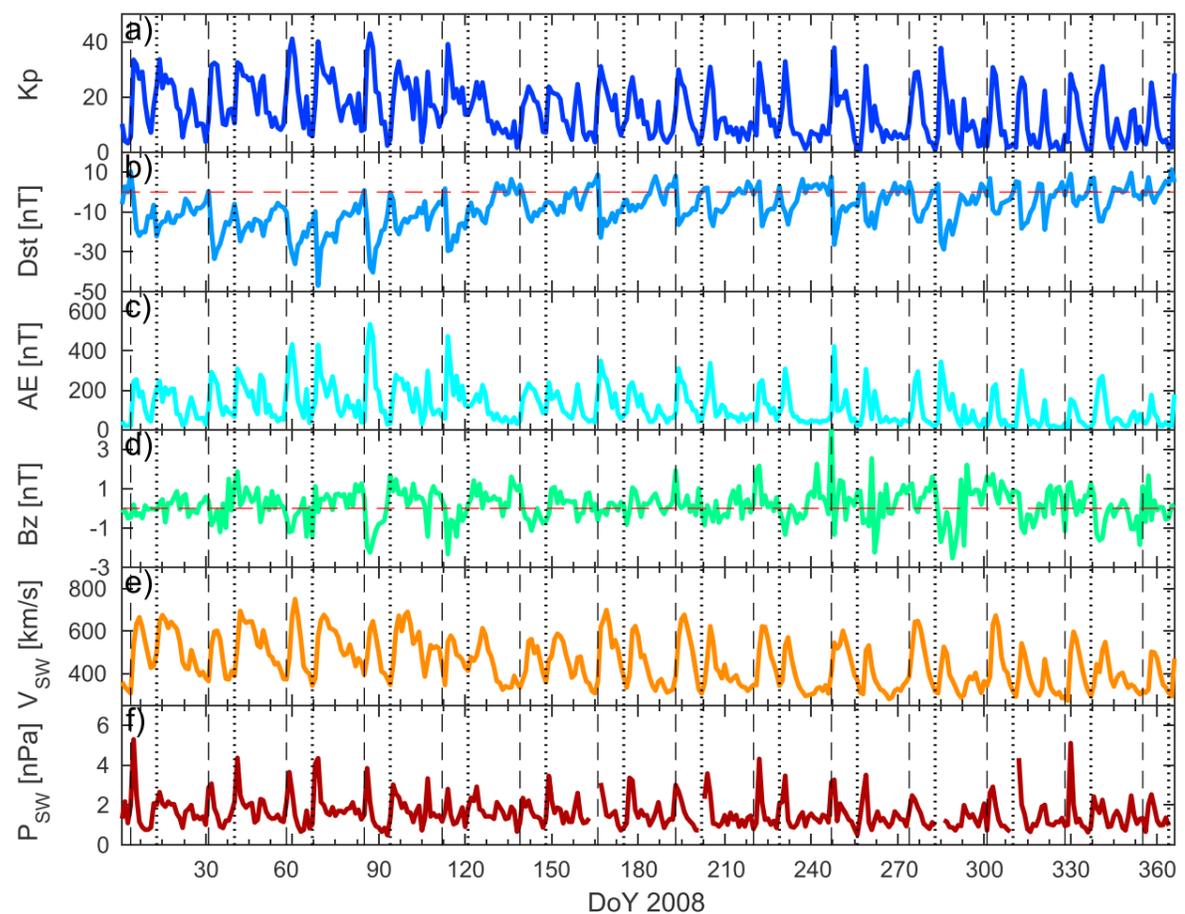

**Figure 2.** Daily means of $Kp$, $Dst$, $AE$, $B_z$, $V_{SW}$, and $P_{SW}$ during 2008 from top to bottom, respectively. The black vertical dashed and dotted lines indicate the 27 days recurrent period of the HSSWS. The red dashed lines show the zero line in the $Dst$ index and $B_z$.

### 3.2. Longitudinal Distribution of OH

To investigate the seasonal longitudinal distribution of OH, we calculate the longitudinal OH running mean (5° longitude window) within a latitude band of 40° to 80° for altitudes 60 to 81 km in both hemispheres for winter, spring, and autumn. Note that due to the SZA selection of >100°, the data coverage for summer stops at about 65°N and 70°S. The months January and February are considered winter in the NH and summer in the SH. While the months June, July, and August are summer in the NH and winter in the SH. The other seasons follow accordingly. However, the month of December 2008 is not included in this study. The seasonal longitudinal distribution of the OH volume mixing ratio (VMR) is shown in Figure 3. All seasons show two OH maxima located at about 67 and 73 km in both hemispheres. Summer months exhibit another OH maximum located at approximately 78 km in both hemispheres. Generally, the OH VMR decreases with decreasing altitude. For winter in the NH, there is high OH VMR within longitudes 150°W–100°E for altitudes 81 km to about 76 km. In the SH winter, the high OH VMR at altitudes 76–81 km is almost homogeneously distributed but strongest within longitudes 180°W–60°W and 120°E–180°E. High OH VMR are still visible in spring within approximately the same longitude region. Autumn also shows high OH VMR within approximately the same longitudes in the NH but weaker compared to winter. In the SH, there is a stronger OH concentration within longitudes 115°–0°W than during winter, with signatures up to below 76 km. Generally, these kind of high OH concentrations at 67–81 km occurred during January, February, March, October, November, and December in the NH, while in the SH, they occurred during April to September (~autumn to winter in both hemispheres).

### 3.3. Longitudinal Variation of EEP Effects on OH

We present the hemispherical distribution of disturbed and quiet conditions for the energy deposition and OH during winter in the NH and autumn in the SH averaged over altitudes between 75 and 78 km in 2008. Disturbed/quiet conditions were sorted based on daily mean energy deposition by EEP at particular altitudes. Days, for which the daily mean energy deposition is greater than the annual mean energy deposition at a particular altitude range, were considered as disturbed time. Whereas days, for which the daily mean energy deposition is less than the annual mean energy deposition at a particular altitude range, were considered to be quiet time. The data are sorted according to geographical latitude and longitude in such a way that each 5° latitude by 10° longitude bin shows the running mean (over a window of three bins) of the energy deposition or OH/temperature/$H_2O$/GPH within that bin.

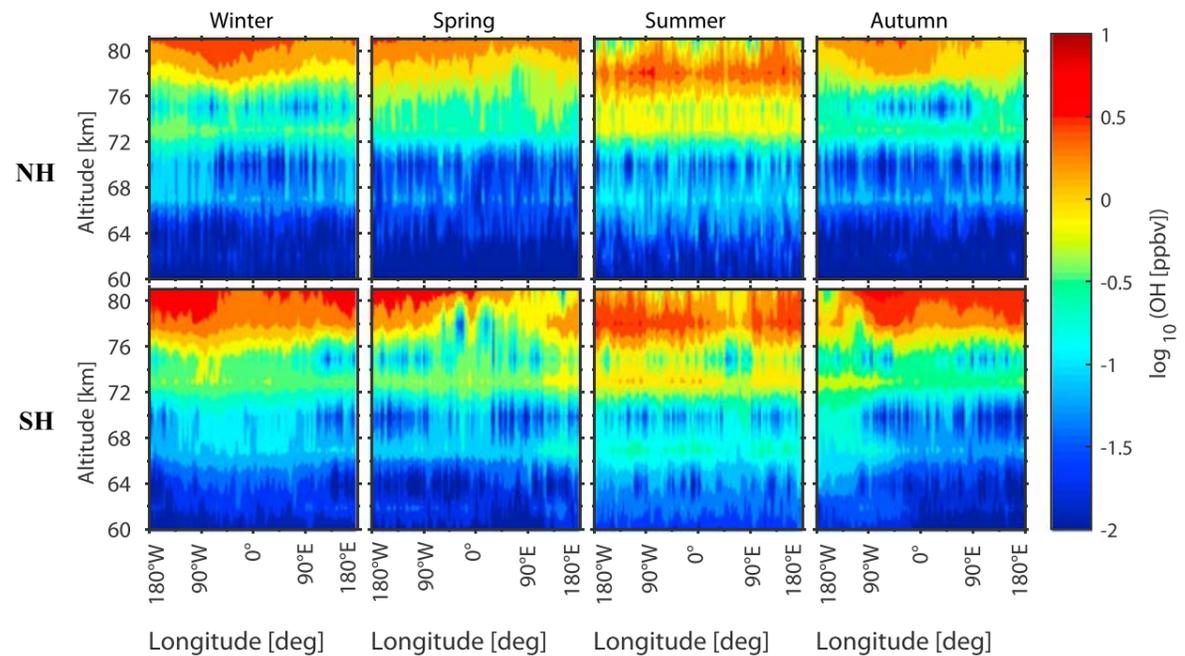

**Figure 3.** OH running mean (5° longitude window) for the altitude range of 60–81 km. Autumn, winter, and spring months cover a latitude band of 40° to 80°N/S for year 2008. Summer months do not extend to 80° latitude due to the SZA selection of >100°. (top row) The NH and (bottom row) the SH. (first column to fourth column) Winter, spring, summer, and autumn. The months January and February are considered winter in the NH and summer in the SH. Whereas the months Jun,e, July and August are considered summer in the NH and winter in the SH. The other seasons follow accordingly.

We present maps of an average at 75–78 km for the energy deposition and OH VMR. (For comparative studies with temperature, $H_2O$ and GPH, we use data at 75 km.) All maps cover geographical latitudes 40°–80° N/S. The first results, sorting by season, show that winter (January and February) in the NH and autumn (March, April, and May) in the SH exhibit the longitudinal signatures in OH reported by *Andersson et al.* [2014b] more clearly than in the other seasons. This is consistent with the strength of the geomagnetic activity. Although winter is the best season to observe EEP effects in the atmospheric OH due to low background levels, EEP-related changes in OH were stronger in the SH autumn compared to winter in 2008. Consequently, we focus on the winter and autumn months for the NH and SH, respectively, as shown in Figures 4 and 5. As the maps of NH and SH cover different months, they also cover different EEP events throughout 2008.

### 3.3.1. Energy Deposition
Figure 4 shows the night-time mean energy deposited between 40° and 80° geographic latitudes during disturbed and quiet conditions for winter and autumn in the NH and SH, respectively. The energy data are averaged between 75 and 78 km. The storm time NH map is fairly homogeneous with maximum values covering northern Russia and part of North America at longitudes (30°E–130°W) and minimum values over Scandinavia to North America (60°W–30°E). In the SH, the energy is deposited almost homogeneously within the latitude range of the radiation belts. The energy deposition during storm times is approximately 1 order of magnitude greater than during nonstorm times in both hemispheres.

### 3.3.2. OH Composition
Figure 5 shows the night-time mean OH maps for disturbed and quiet conditions during winter in the NH and autumn in the SH. In the NH, OH shows clear enhancements poleward of latitude 55°N CGM during storm conditions, with local maxima within longitudes 90°–10°W and 70°E–130°W which *Andersson et al.* [2014b] refers to as the North America and northern Russia hot spots. The OH enhancement over North America is, however, also present during quiet times within longitudes 90°–0°W, and it is stronger over the North Atlantic Ocean than over North America. (There is also a region of high-OH volume mixing ratio at 60–90°E about 40–45°N ($L \sim 1.5-2$) corresponding to the inner radiation belt. As our focus is the auroral and subauroral latitudes, we consider this feature outside the scope of the current paper.)

In the SH, there appears to be a local OH enhancement over West Antarctica both during disturbed and quiet conditions which *Andersson et al.* [2014b] refers to as the Antarctic Peninsula hot spot. OH enhancement is seen at all longitudes during disturbed times and within longitudes 150°–0°W during quiet conditions. The West Antarctica hot spot appears, however, to be unbound by the geomagnetic location of Earth's radiation belts but seems rather to be more geographically constrained.

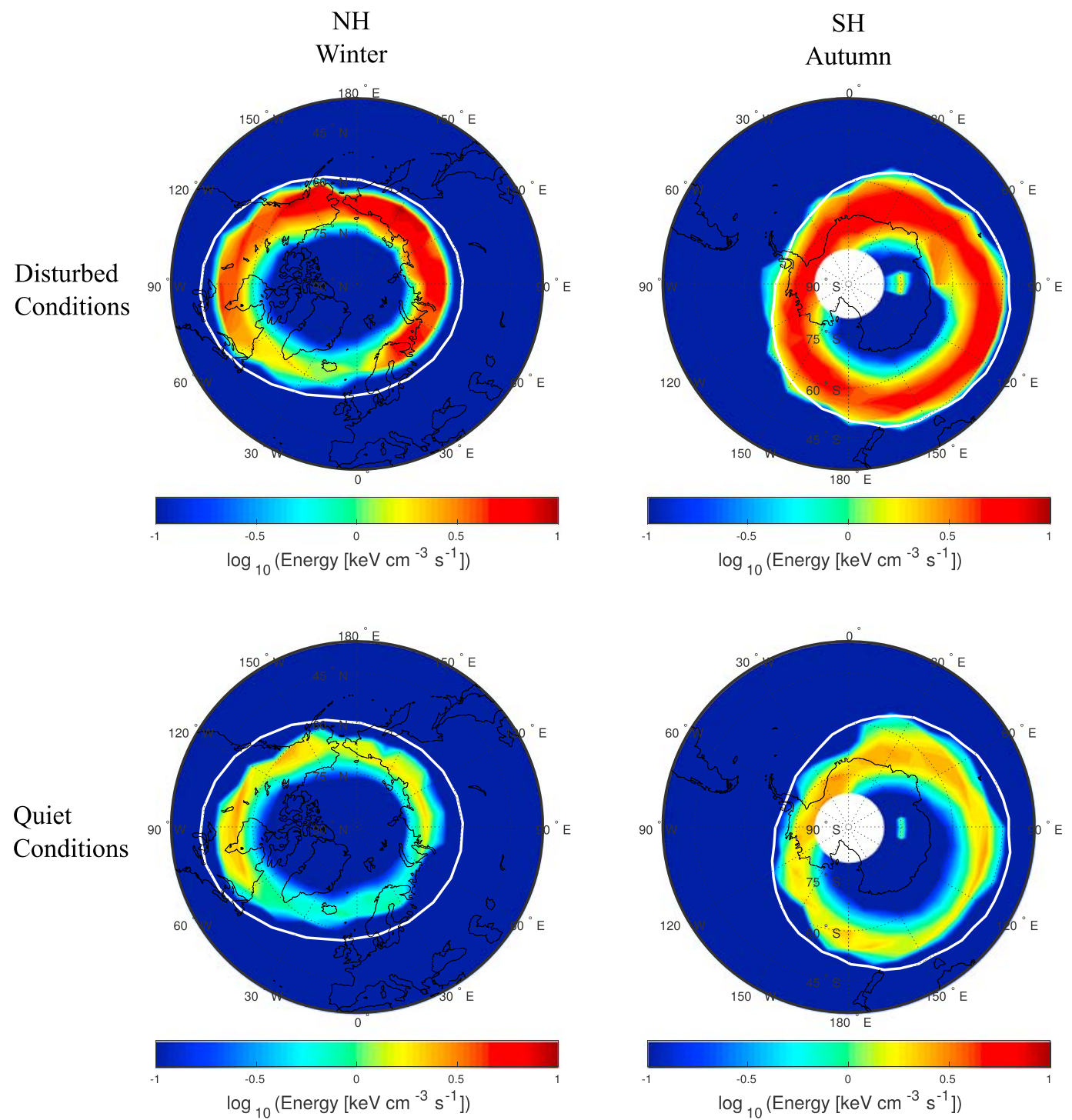

**Figure 4.** Mean nighttime energy deposition during (top row) disturbed and (bottom row) quiet conditions at altitudes 75–78 km for (left column) winter (January–February) at LST $2 \leq LST < 4$ in the NH and for (right column) autumn (March–May) at LST $22 \leq LST \leq 1$ in the SH. Mean values were calculated for each 5° latitude by 10° longitude bin between 40° to 80°N and longitudes 180°W to 180°E. The white line shows the approximate location of 55°N/S CGM latitude.

Note that the local OH enhancements in Figure 5 generally cover smaller longitude ranges than the hot spots seen by *Andersson et al.* [2014b] but are located within the same regions. Both the current presentation and the maps presented by *Andersson et al.* [2014b] show OH features (signatures) unconstrained to the geomagnetic latitude location (or footprints) of the Earth's radiation belts, indicating potential signatures due to the background dynamics.

### 3.4. Dynamical Background

We investigate the possibility that atmospheric dynamics is responsible for some of the observed OH enhancements. Figure 6 shows the temperature, $H_2O$ and GPH (top row to bottom row) for quiet time conditions in the NH (left column) and SH (right column), respectively. In the NH, there is a temperature enhancement between longitudes 90°W and 110°E and an $H_2O$ minimum. The temperature maximizes in approximately the same region where the North America OH maximum is located. A depression in the GPH is seen within longitudes 180°W–0°W, displaced by approximately 90° westward from the location of the temperature maximum and $H_2O$ minimum. In the SH, the same features are seen in the temperature, $H_2O$, and GPH for the West Antarctica maximum, all located over approximately the same region over West Antarctica without a shift in the location of the GPH minimum. The North America and West Antarctica OH enhancements seem to follow more closely the region of intersection of the depression in the GPH with temperature maximum and $H_2O$ minimum.

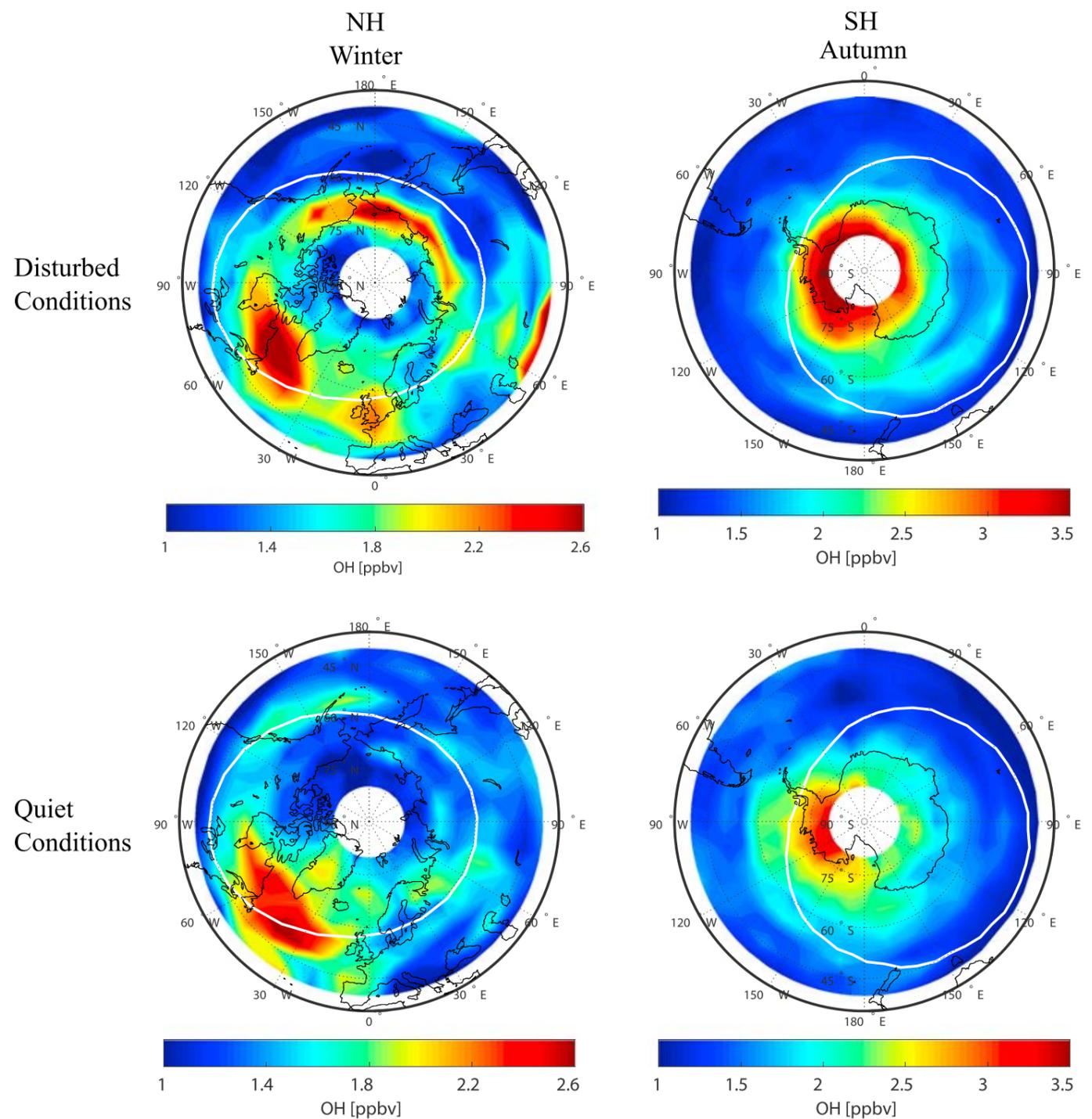

**Figure 5.** Mean nighttime OH during (top row) disturbed and (bottom row) quiet conditions at altitudes 75–78 km for (left column) winter (January–February) at LST 2 ≤ LST < 4 in the NH and for (right column) autumn (March–May) at LST 22 ≤ LST < 2 in the SH. Mean values were calculated for each 5° latitude by 10° longitude bin between 40° to 80°N and longitudes 180°W to 180°E. The white line shows the approximate location of 55°N/S CGM latitude.

During winter time, especially in the NH, planetary wave activity is known to play an essential role in the background dynamics. To determine the role of planetary wave activity upon the OH composition, Figure 7 shows the OH (a), EEP (b), and temperature anomaly (c) in a geographic latitude band 60°–70°N for quiet (right column) and disturbed (left column) conditions. We sorted the data based on the energy deposition as in section 3.3. The temperature anomaly is the difference from the mean over the period. Then we derive the quasi-stationary planetary wave numbers 1 (S1) and 2 (S2) shown as the superposition of two sinusoidal curves fitted to the longitudinal temperature anomaly Figures 7c and 7d.

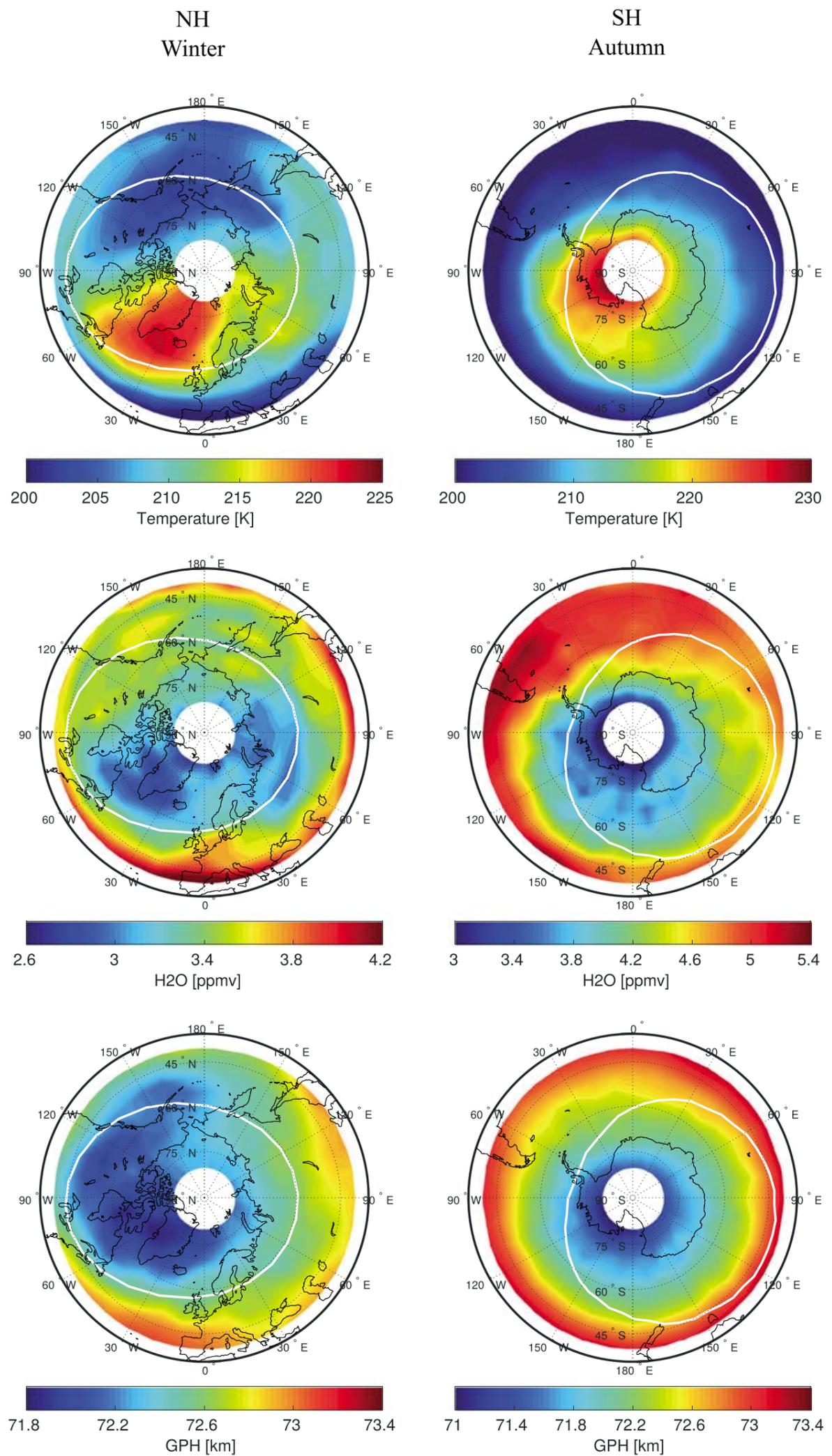

**Figure 6.** (top row to bottom row) Mean nighttime temperature, H₂O, and GPH during quiet conditions at altitudes 75 km for (left column) winter (January–February) at 2 ≤ LST < 4 in the NH and for (right column) autumn at LST 22≤LST<2 in the SH during 2008. Mean values were calculated for each 5° latitude by 10° longitude bin between 40° to 80°S and longitudes 180°W to 180°E. The white line shows the approximate location of 55°N/S CGM latitude.

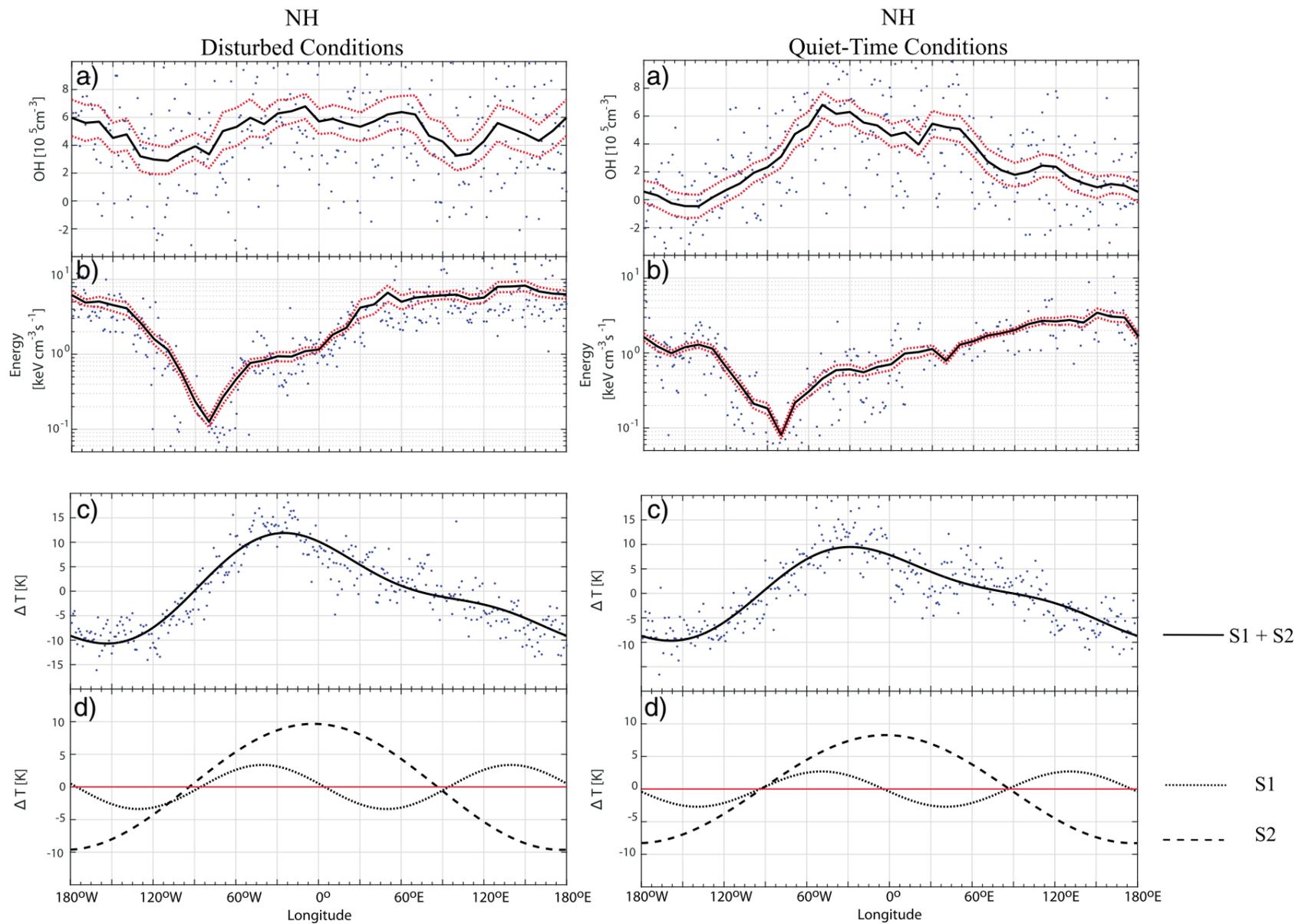

**Figure 7.** Longitudinal variation of OH and energy deposition averaged between altitudes 75–78 km during winter for the latitude band of 60–70°N in 2008 for disturbed and quiet time conditions. For comparison with background atmospheric dynamics, Figures 7c and 7d show the quasi-stationary planetary wave activity for wave number 1 and 2 derived from temperature between latitudes 60 and 70°N at an altitude of 75 km during winter for year 2008. (top row to bottom row) (a) OH (blue). Running mean of 30° longitude (black). (b) Energy deposition (blue). Running mean of 30° longitude (black). (c) Temperature anomaly (blue) fitted with a sinusoidal curve for the superposition of planetary waves S1 and S2 (black). (d) Sinusoidal curve fitting for planetary wave numbers 1 (S1) and 2 (S2). The red solid line is the zero line. The red dotted lines show the standard error of the mean.

During NH winter in 2008 the S1 and S2 peak around the same longitudes during both quiet and disturbed conditions. S1 has its maximum around 0°E. S2 has its maximum around 140°E and 40°W. The amplitudes are slightly higher during disturbed conditions. However, this may well be due to the random sample of days for the storm time conditions and unrelated to the EEP. During disturbed conditions S1 and S2 have an amplitude of about 19 K and 7 K, respectively. The superposition of the two waves, S1 + S2 is large at 75 km driving a longitudinal variability in temperature of around 23 K with a maximum around 25°W and a minimum close to 155°W.

The OH during disturbed conditions shows that in addition to changes correlated with the temperature variations, an additional source is present that coincides with the distribution of EEP. As Figure 4 shows, the EEP in a geographic latitude band between 60°–70°N samples only part of the auroral oval due to the offset between geomagnetic and geographical coordinates. The energy deposition shows high values (>4 keV $cm^{-3}s^{-1}$) between longitudes 25°E and 135°W corresponding well with the increase of OH in this region. In the region of little EEP, the OH appears to track the temperature variations as it does during quiet times. To better reveal the EEP impact on the OH density, we have subtracted the quiet time conditions from the disturbed conditions as shown in Figure 8. The longitudinal OH behavior is generally in phase with the energy deposition. The only exception is the longitude interval 30°–60°E which corresponds to regions where the auroral oval intersects a descending area of low $H_2O$ mixing ratios (see Figure 6). Note that the negative OH values are due to the fact that some of the MLS observations are noisy in nature. Ignoring such values will automatically introduce a positive bias into any averages made of the data as part of scientific analysis [*Livesey et al.*, 2015].

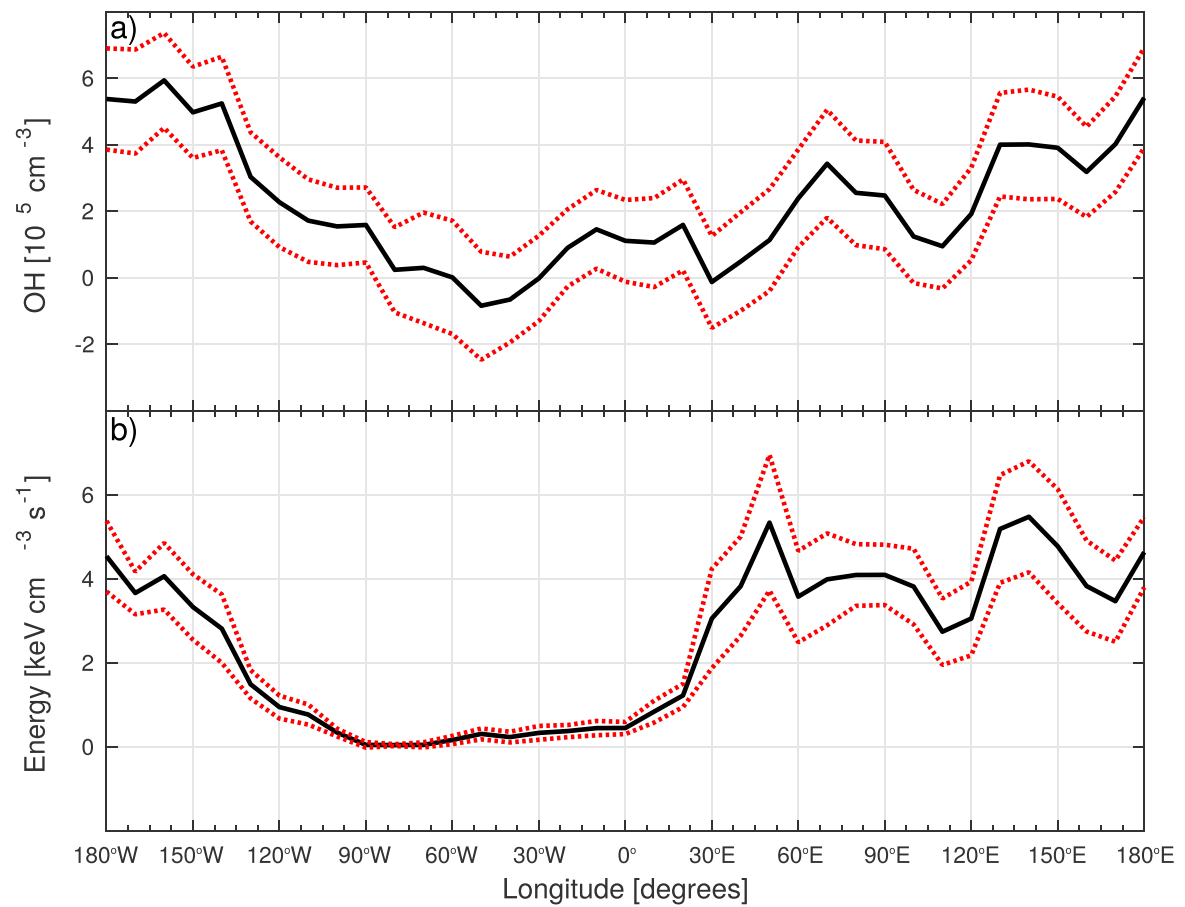

**Figure 8.** The difference between disturbed and quiet time at 75–78 km within a latitude band of 60–70°N for winter in 2008. The red dotted lines show the standard error of the mean. (a) Longitudinal OH (b) longitudinal energy deposition.

## 4. Discussion

### 4.1. Is the Impact of EEP Upon the OH Production of Any Significance During Solar Minimum?

During geomagnetic storms radiation belt particles can be accelerated to high and possibly relativistic energies that precipitate deep into the atmosphere, causing enhancements of OH. EPP leads to production of $HO_X$ species through ionization, dissociation, and dissociative ionization of the most abundant chemical species in the atmosphere ($N_2$ and $O_2$). The abundance of $H_2O$ below 80 km facilitates the formation of large water cluster ions which recombine with electrons forming ~2$HO_X$ per ionization [*Solomon et al.*, 1981; *Sinnhuber et al.*, 2012]. EEP appears to impact the geomagnetic latitudes 55°–72° N/S CGM as illustrated by Figure 4 [see also *Andersson et al.*, 2014b].

The region of high OH VMR over northern Russia is collocated with the region of enhanced EEP energy deposition. The region of high OH VMR over North America and the North Atlantic Ocean, prominent during both disturbed and quiet conditions is not evident in the energy deposition. In the SH, we find a similar situation. The OH concentration maximizes over the West Antarctica, while the energy deposition is found to be rather homogeneously distributed with longitude. In the following, we will discuss the extent to which the local OH enhancements in the two hemispheres are related to the EEP energy deposition and the role of the background atmosphere on the longitudinal distribution of OH.

#### 4.1.1. Longitudinal Variations of the Energy Deposition and OH in the NH

For January 2005 to December 2009, *Andersson et al.* [2014b] found two regions of high OH concentration during high EEP in the NH (51 days of data in total). These regions are the North America and northern Russia, which were attributed to EEP forcing. Our energy deposition maps based on the electron fluxes from MEPED/NOAA 18 maximizes as illustrated in Figure 4 over the northern Russia region. We do not, however, find evidence based on the energy deposition for an EEP-produced OH concentration enhancement over North America. The OH enhancement is prominent in both disturbed and quiet times. The energy deposition pattern cannot explain the distinct pattern found in the OH concentration.

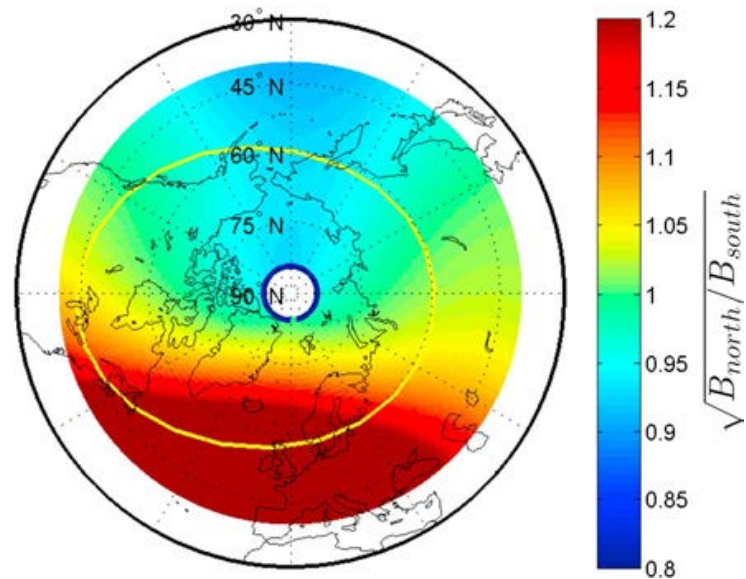

**Figure 9.** Ratio of magnetic field strength in the NH to the magnetic field in the SH found by magnetic field tracing using the IGRF model.

In addition to covering a shorter time period, some of the discrepancies between our energy deposition map and the map provided by *Andersson et al.* [2014b] might be due to the fact that they use the 0° detector E1 channel measuring electrons of energy larger than 30 keV. The E1 channel counts might be dominated by the lower energies that will not penetrate below 80 km. As we estimate the energy deposition at the respective heights we use the information from all the energy channels. We also limit the electron flux analysis to NOAA/POES 18 which is close in both time and space to the Aura OH retrieval. *Andersson et al.* [2014b] used multiple spacecraft at different MLT regions compared to the MLT region of the OH retrieval from Aura.

*Barth et al.* [2001] investigated the geomagnetic longitude dependence observed in $NO_X$ at 106 km produced by auroral electrons. The most characteristics feature in their result is a minimum in a region above Scandinavia and Greenland. A possible candidate for these longitudinal variations in the NH is the asymmetries in the Earth's magnetic field. Electrons drifting around the Earth to the weak field associated with the SAMA are lost to the atmosphere. This depletes the electrons throughout the anomaly region. *Barth et al.* [2001], therefore, suggested that we are seeing the "normal" electron precipitation west of the weak magnetic field region but a much weakened precipitation of electrons within the weak magnetic field region itself and eastward of it. Figure 9 shows the square root ot the ratio: $B_{North}/B_{South}$, where $B_{South}$ is found by magnetic field tracing using the IGRF model. Based on this theory, the low magnetic field ratio over North Asia and Alaska would imply that there will be more electron precipitation there than in the SH for particles bouncing over the same magnetic field lines. Therefore, the lower magnetic field ratio and the associated electron precipitation may explain the presence of the northern Russia hot spot. On the other hand, the magnetic field ratio is relatively higher above Scandinavia and Greenland. As the high $B_{North}/B_{South}$ ratio overlaps with the large OH VMR, it does seem unlikely that the North America/North Atlantic OH maximum is due to EEP forcing.

### 4.1.2. Longitudinal Variations of the Energy Deposition and OH in the SH

In the SH autumn, there is a maximum in the OH density above West Antarctica during both disturbed and quiet conditions. The West Antarctica OH enhancement and its persistence during quiet times might be explained by the weaker magnetic field in this region that allows a steady drizzle of radiation belt electrons. Electrons that were mirroring at other longitudes could be lost here as they penetrate deeper in the atmosphere and interact with the denser atmosphere. The West Antarctica hot spot is also visible in winter and spring starting at approximately 70–73 km altitude during quiet times (not shown).

According to *Horne et al.* [2009], the effects on atmospheric chemistry due to relativistic electron precipitation (REP) are more likely to occur in the SH poleward of the SAMA region, because >1 MeV electron precipitation occur mainly in that region. However, the energy deposition seems rather uniformly distributed with geographic longitude circles as seen in Figure 4. This is also a feature in the results of *Andersson et al.* [2014b].

Although the West Antarctica OH enhancement can be explained by the weaker magnetic field in that region, its features seem more constrained by geographical rather than geomagnetic location. There is a possibility that in addition to the above mentioned causes, atmospheric dynamics may also play an active role in the formation of the West Antarctica OH maximum.

### 4.2. The Role of the Background Dynamics in Determining the Longitudinal OH Distribution

Elevated temperatures and dry air at 75 km appear to coexist with the longitudinal region of elevated OH VMR measured by Aura as shown in Figure 6. The high temperatures are associated with descending air motions, bringing down dry air from higher altitudes. The descent will then also bring down odd oxygen (O and $O_3$) and atomic hydrogen (H). Atomic oxygen has a large concentration gradient from the middle mesosphere to the mesopause, which makes the atomic oxygen highly variable in the presence of vertical motion. Even a small displacement can generate large changes in the mixing ratio [*Smith*, 2004]. Thus, the $O_3$ mixing ratio at 75 and 78 km will increase if $O_3$-rich air from the secondary $O_3$ maximum is brought down.

*Winick et al.* [2009] found elevated OH Meinel emissions related to the vertical displacements associated with SSWs. The vertical displacement of the OH airglow layer from 87 km to 78 km was observed at the time of major SSW from January 2009 [*Shepherd et al.*, 2010]. Assuming that this transport is fast enough to maintain an O density, it could lead to an enhanced production of vibrationally excited OH (OH*) at lower than normal altitudes [*Winick et al.*, 2009]. The pressure here will be sufficient to allow a third body reaction creating $O_3$:

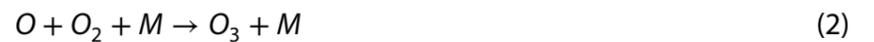

$$O + O_2 + M \rightarrow O_3 + M \qquad (2)$$

Then, $O_3$ will react with atomic oxygen forming OH by the exothermic reaction:

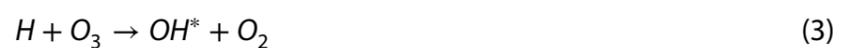

$$H + O_3 \rightarrow OH^* + O_2 \qquad (3)$$

The variability of the ground state OH measured with the Aura MLS instrument during periods of SSWs by *Damiani et al.* [2010] further corroborates the interpretation by *Winick et al.* [2009]. The OH* is deactivated either by photon emissions in the Meinel band (observed in the airglow) or by collisional quenching [*Brasseur and Solomon*, 2005]. The latter depends on the density and therefore becomes more important at lower altitudes. The enhanced level of atomic hydrogen might also contribute to the conversion of $HO_2$ into OH by the reaction [*Brasseur and Solomon*, 2005]:

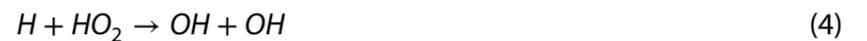

$$H + HO_2 \rightarrow OH + OH \qquad (4)$$

This potential chemical scheme seems consistent with our observation. As pointed out earlier, the quiet time longitudinal OH enhancement is only apparent above 73 km, supporting the potential effect of a steep gradient in the mixing ratios of odd oxygen. In January and February 2008 the polar vortex was displaced from its climatological position over the pole by an SSW [*Medvedeva et al.*, 2012]. There could therefore be a systematic longitudinal response associated with SSW also in our observations. However, considering the complex chemistry of the odd hydrogen family and its dependence on temperature, pressure, and mixing ratios, modeling studies and/or additional satellite data are needed in order to quantitatively assess the described features.

Also, the $H_2O$ mixing ratios might impact the OH production efficiency. At altitudes above about 65 km, $HO_x$ production depends on the ionization rate, and the atomic oxygen and $H_2O$ densities [*Solomon et al.*, 1981]. The main process during the formation of $HO_x$ due to EPP events is the uptake of $H_2O$, forming large cluster ions and subsequent recombination with electrons. However, if the $H_2O$ mixing ratios are reduced by a few ppb (parts per billion), water cluster reactions may be cut off by dissociative recombination of the intermediates (e.g., $O_4^+$, $O^+{}_2$ $H_2O$) with electrons. In this case, even the natural electron concentration may be sufficient to reduce the efficiency of the $HO_x$ production rate [*Crutzen and Solomon*, 1980]. This implies that the impact of EEP might also depend directly on the dynamical background as the EEP deposited over North America may be less efficient in producing OH due to the low $H_2O$ mixing ratios compared to the EEP over northern Russia which has higher $H_2O$ mixing ratios. This explanation also applies to Figure 8 for longitudes 30°–60°E, and it is briefly mentioned in subsection 3.4.

Planetary waves are more prominent during wintertime, in particular in the NH. This is evident in Figure 7 which shows that the general longitudinal trend of OH VMR within a latitude band of 60°–70°N closely follows that of the superposition of the planetary waves S1 + S2 especially during quiet time conditions. Quasi stationary planetary wave activity is of approximately the same amplitude and phase during both disturbed and quiet times (Figures 7c and 7d). Quasi-stationary planetary waves drive the background OH density with a peak that persists during both disturbed and quiet time conditions regardless of the strength of the energy deposition. The North America OH maximum which is present during both disturbed and quiet time conditions is a feature attributed to quasi-stationary planetary wave activity.

Any EEP-induced OH production will be an addition onto the already existing background OH. The OH enhancement due to EPP is visible for strong energy deposition during disturbed conditions. The visibility of the OH variability due to EEP depends on the background OH and the strength of the energy deposited. Hence, during disturbed conditions, there exist two peaks in the OH density: one due to the background dynamics and another due to EPP. Figure 7 does not show an exact one to one relationship between the energy deposition and the OH density in regions (longitudes 25°E–135°W) of high energy deposition where such a relationship may be expected. As already discussed, the OH production efficiency is affected by the $H_2O$ density which may be variable over the latitude band under consideration.

Therefore, we believe that in our results for the winter 2008 NH, it is only the OH maximum over northern Russia that is attributed to EEP forcing, while the North America hot spot is mainly a consequence of a dynamical atmosphere. The West Antarctica hot spot in the SH is attributed to both EEP and atmospheric dynamics. This mechanism has not been considered by *Andersson et al.* [2014b]. Based on the increased level of OH they find over Greenland, which does not seem to be restricted in geomagnetic latitude, a planetary wave effect might be present in their data as well. Although our analysis supports the conclusion that even small storms can impact the mesospheric OH, a quantitative assessment needs to firmly establish the dynamically varying background. This could potentially be achieved by applying a multilinear regression analysis of OH with GPH, $O_3$, and H. The production of OH due to EEP could then be included with a potential dependency on $H_2O$.

## 5. Summary and Conclusions

OH enhancements due to EEP were seen poleward of CGM latitudes 55°N/S with regions of local maxima: the northern Russia and West Antarctica maxima. We find that the West Antarctica maximum might be explained by a combination of EEP and the weak magnetic field in this region. Even in the cases with geomagnetic quiet conditions, the weaker magnetic field in this region causes a steady drizzle of radiation belt electrons, hence causing an OH "pool" in this region. In addition, atmospheric dynamics might contribute to the formation of the West Antarctica maximum.

The North America OH maximum cannot be explained by the weaker magnetic field, since it is located in a region with relatively high magnetic field ratio. It rather appears that the North America OH maximum in 2008 is due to dynamical effects. Planetary waves in the polar winter induce downwelling of thermospheric air, bringing down dry air, atomic oxygen, and atomic hydrogen. The air density at mesospheric altitudes is sufficient to facilitate three-body reactions between atomic oxygen and $O_2$, which results in formation of $O_3$, which again reacts with atomic hydrogen forming OH. The same dynamical features may be related to the West Antarctica OH maximum. The northern Russia OH maximum, however, appears to be a feature related to EEP forcing alone.

Our findings suggest that even during solar minimum, there is substantial EEP driven OH production. To quantify this effect, the background atmospheric dynamics have to be taken into account, along with detailed knowledge of where and when the precipitation occurs. Background atmospheric dynamics are important in explaining the longitudinal distribution of OH.

**Acknowledgments**
This study was supported by the Research Council of Norway under contract 223252/F50. The authors thank the NOAA's National Geophysical Data Center (NGDS) for providing NOAA data (http://satdat.ngdc.noaa.gov/), WDC Geomagnetism, Kyoto, Japan, for *AE* and *Dst* indices (http://wdc.kugi.kyoto-u.ac.jp/wdc/Sec3.html), SPDF Goddard Space Flight Center for solar wind parameters (http://omniweb.gsfc.nasa.gov/), and NASA Goddard Earth Science Data and Information Services Center (GES DISC) for providing Aura/MLS data (http://mls.jpl.nasa.gov/).